\documentclass[twocolumn,showpacs,showkeys,amsmath,amssymb,preprintnumbers,nofootinbib,floatfix,prl]{revtex4}

\usepackage{graphicx}% Include figure files
\usepackage{bm}% bold math
\usepackage{mathrsfs}
\usepackage{amsmath}
\usepackage{verbatim}
\usepackage[all, knot]{xy}
\xyoption{arc}
\xyoption{knot}
\xyoption{2cell}
\xyoption{poly}
\xyoption{graph}

\begin{document}

\preprint{Version 1.0}

\title{Quantum logic as superbraids of entangled qubit world lines}

\author{Jeffrey Yepez}

\address{${}^{1}$Air Force Research Laboratory, Hanscom Air Force Base, Massachusetts  01731}

\begin{abstract}
 Presented is a topological representation of quantum logic that views entangled qubit spacetime histories (or qubit world lines) as a generalized braid, referred to as a superbraid. The crossing of world lines is purely quantum in nature, most conveniently expressed analytically with ladder-operator-based quantum gates.  At a crossing, independent world lines can become entangled.  Complicated superbraids are systematically reduced by recursively applying novel quantum skein relations.  If the superbraid is closed ({\it e.g.} representing quantum circuits with closed-loop feedback, quantum lattice gas algorithms, loop or vacuum diagrams in quantum field theory), then one can decompose the resulting superlink into an entangled superposition of classical links.  In turn, for each member link, one can compute a link invariant, {\it e.g.} the Jones polynomial.  Thus, a superlink possesses a unique link invariant expressed as an entangled superposition of classical link invariants.  
\end{abstract}

% PACS, the Physics and Astronomy Classification Scheme.


\maketitle

In topological quantum computing \cite{PhysRevLett.90.067902,nayak:1083}, a quantum gate operation derives from braiding quasiparticles, {\it e.g.} two Majoranna zero-energy vortices made of entangled Cooper-pair states in a $p + ip$ superconductor  where the vortex-vortex phase interaction has a non-abelian SU(2) gauge group  \cite{PhysRevLett.86.268,tewari:010506}.    Dynamically braiding such quantum vortices (point defects in a planar cross-section of the condensate)  induces phase shifts in the quantum fluid's multiconnected wave function. Local nonlinear interactions between deflects (vortex-vortex straining) is otherwise neglected, {\it i.e.} the separation distance  $\delta$ of the zero mode vortices is much greater then the vortex core size, which scales as the coherence length $\xi \ll \delta$  in quantum fluids.  The braiding occurs adiabatically so the quantum fluid remains in local equilibrium and the number of deflects (qubits) remains fixed.   For implementations, the usual question is how can quantum logic gates, and in turn quantum algorithms, be represented by braiding deflects, quasiparticles with a nonabelian gauge group.  

This Rapid Communication addresses the related fundamental question 
about the relationship between quantum entanglement, tangled strands, and quantum logic  \cite{Kauffman:2004p1715}.  How can  a quantum logic gate, and in turn a quantum algorithm, be decomposed into a linear combination (entangled superposition) of classical braid operators?  The goal is
to comprehend and categorize quantum algorithms topologically.   This is done by first viewing  a quantum gate as  a braid of two qubit spacetime histories or world lines.  Qubit-qubit interaction associated with a quantum gate is rendered as a tree-level scattering diagram, a form of  ribbon graph.  Then, a quantum algorithm is represented as a weave of such graphs, a {\bf superbraid} of qubit world lines.  Finally, one closes a superbraid to form a {\bf superlink}.  In fact, quantum lattice gas algorithms, {\it e.g.}  employed for the simulation of superfluids themselves \cite{yepez-vahala-arXiv-0905.0159v1}, are a good superlink archetype, hence the shared nomenclature.  

With this technology we can calculate superlink invariants (Laurent series in linear combination). 
In principle, each algorithm has its own unique linear combination of invariant Laurent polynomials; {\it e.g.} two competing quantum circuit implementations of a particular algorithm can be judged equivalent, irrespective of circuit schemes and number of gates and wires.  
If two quantum algorithms, first and second-order accurate, are topologically equivalent, then the simpler one can be used for analytical predictions of their common effective theory while the latter for faster simulations on smaller grids.

In short, presented is a quantum generalization of the Temperley-Lieb algebra TL${}_Q$ and Artin braid group $B_Q$:  a superbraid and its closure, a superlink, is formed out of the world lines of $Q$ qubits (strands) undergoing dynamics generated by quantum gates.   Furthermore,  the superbraid representation of quantum dynamics works equally well for either bosonic or fermionic quantum simulations.   There exists a classical limit where the quantum Temperley-Lieb algebra and the superbraid group, defined below, reduce to the usual Temperley-Lieb algebra and braid group.

Classical braid operators (nearest neighbor permutations), represented in terms of Temperley-Lieb algebra \cite{Temperley04201971}, were originally discovered in six-vertex Potts models and statistical mechanical treatments of two-dimensional lattice systems \cite{Baxter.1982,PhysRevLett.64.499}.  The quantum algorithm to compute the Jones polynomial  \cite{jones1132579,kauffman:65730T} employs unitary gate operators that are mapped to unitary representations of the braid group, {\it i.e.} generated by hermitian representations of the Temperley-Lieb algebra.  
To prepare for our presentation of superbraids as a novel topological representation of the quantum logic underlying quantum information dynamics, let us first briefly mention some basics of knot theory and some basic quantum gate technology using qubit ladder operators.

A link comprising $Q$ strands, denoted by $L$ say, is the closure of a braid.  The Jones polynomial $V_L(A)$ is an invariant of $L$ \cite{Jones:1985p1577}, where $A$ is a complex parameter associated with the link whose physical interpretation will be presented below.  $V_L(A)$ is a Laurent series in $A$. The Jones polynomial is defined for a link embedded in three space--an oriented link. One projects $L$ onto a plane.  In the projected image, in general crossing of strands occurs but is disambiguated by its sign $\pm 1$, {\it i.e.}  one assigns over-crossings the sign of  $+1$ and under-crossings $-1$.   The writhe $w(L)$ is sum of the signs of all the crossings, {\it i.e.} the net sign of a link's planar projection.  
The Jones polynomial is computed as follows
\begin{equation}
V_{L}(A) = \frac{1}{d}\,(-A^{3})^{w(L)}
 \,K_{L}(A),
\end{equation}
where $K_L(A)$ is  the Kauffman bracket of the link.   $K_L(A)$  is determined by summing over all possible planar projections of $L$.   In the simplest case of an unknotted link (or unknot), the Kauffman bracket is
\begin{equation}
\label{unknot_rule}
\xy
(-9,0)*{\cir<8pt>{}};
(17,0)*{=K_\bigcirc(A) = d= - A^2  - A^{-2}.};
\endxy
\end{equation}
The Kauffman bracket of a disjoint union of $n$ unknots has the value $d^n$, {\it e.g.} 
\(
\xy
(-6,0)*{\cir<6pt>{}};
(0,0)*{\cir<6pt>{}};
(6,0)*{\ = d^2.};
\endxy
\)

$K_L(A)$  for a link with crossings can be computed recursively using a skein relation that equates it
to the weighted sum of two links, each 
with one less crossing:
\begin{subequations}
\label{skein_relations}
\begin{equation}
\label{hcross_neg_rule}
\xy 
0;/r1.5pc/:
(0,0.25)*{\hcrossneg};
(2,0)*{= A};
(1.3,0.25)*{\huntwist};
(5,0)*{+ \ A^{-1}};
(3,0.25)*{\huncross};
\endxy 
\end{equation}
\begin{equation}
\label{hcross_rule}
\xy 
0;/r1.5pc/:
(0,0.25)*{\hcross};
(2,0)*{= A};
(1.3,0.25)*{\huncross};
(5,0)*{+ \ A^{-1}};
(3,0.25)*{\huntwist };
\endxy 
\end{equation}
\end{subequations}
where $A$ and its inverse are the weighting factors.
As an example, let us recursively apply (\ref{skein_relations}) to prove an intuitively obvious link identity 
$
\xy
0;/r1pc/:
(1,0.25)*{\hcap-\hcrossneg\hcross\hcap};
(5,0)*{=};
(3,0.25)*{\hcap-\hcap};
\endxy
\ \, \quad
$. 
 One reduces the relevant braid as follows
\begin{subequations}
\label{braid_identity_1}
\begin{eqnarray}
% (a)
\xy
0;/r0.75pc/:
{};
(-2,0.615)*{\hcrossneg\hcross};
\endxy
&\stackrel{(\ref{hcross_neg_rule})}{=}&
\xy
0;/r0.75pc/:
(1,0.25)*{ A};
(1,0.365)*{\huntwist \hcross};
(6,0.25)*{ + \ A^{-1}};
(4,0.365)*{\hunover\hcross};
\endxy
% (b)
\\
&\stackrel{(\ref{hcross_rule})}{=}&
\xy
0;/r0.75pc/:
(3,0.25)*{A^2};
(2,0.365)*{\huntwist \huncross};
(7,0.25)*{ + };
(4.0,0.365)*{\huntwist \huntwist};
+(5,0)*{+ \ A^{-1}};
(7,0.365)*{\hunover\hcross};
\endxy
\\
% (c)
&\stackrel{(\ref{hcross_rule})}{=}&
\xy
0;/r0.75pc/:
(3,0.25)*{A^2};
(2,0.365)*{\huntwist \huncross};
(7,0.25)*{ + };
(4.0,0.365)*{\huntwist \huntwist};
+(4,0)*{+};
(6,0.365)*{\hunover\huncross};
+(5,0)*{+ \ A^{-2}};
(9,0.365)*{\hunover \huntwist};
\endxy
\qquad
\\
% (d)
&\stackrel{(\ref{unknot_rule})}{=}&
\xy
0;/r0.75pc/:
(3,0.25)*{A^2};
(2,0.365)*{\huntwist \huncross};
(7,0.25)*{ + };
(4.0,0.365)*{\huntwist \huntwist};
+(4,0)*{\ + \ d};
(6.25,0.365)*{\huncross};
+(5,0)*{+ \ A^{-2}};
(9,0.365)*{\hunover \huntwist};
\endxy
\\
% (e)
&=&
\xy
0;/r0.75pc/:
(0.5,0.365)*{\huntwist};
(7,0.25)*{+\left(d + A^2 + A^{-2}\right)};
(5.75,0.365)*{\huncross};
\endxy
\\
% (f)
&\stackrel{(\ref{unknot_rule})}{=}&
\xy
0;/r0.75pc/:
(0.5,0.625)*{\huntwist};
\endxy
\end{eqnarray}
\end{subequations}

A quantum gate represents the qubit-qubit coupling that occurs at the crossing of world lines of a  pair of qubits,  say $|q_\alpha\rangle$ and $|q_\gamma\rangle$ in a system of $Q$ qubits. Every quantum gate is generated by 
an hermitian operator, ${\cal E}_{\alpha\gamma}$ say, and whose action on the quantum state may be expressed as
\begin{equation}
\label{basic_conservative_quantum_gate}
|\dots q_\alpha \dots q_ \gamma\dots \rangle' = e^{i \zeta {\cal E}_{\alpha \gamma} } |\dots q_\alpha\dots q_ \gamma\dots\rangle,
\end{equation}
where $\zeta$ is a real parameter.
 The archetypal case considered here is ${\cal E}_{\alpha \gamma}^2= {\cal E}_{\alpha \gamma}$; the generator is idempotent.

 Suppose the system of qubits is employed to model the quantum dynamics of fermions or bosons. Is there an analytical form of the generator ${\cal E}_{\alpha\gamma}$ that allows one to easily distinguish between the two cases?   It is natural to begin by treating fermion statistics.  With the logical one state of a qubit 
\(
{\scriptsize
|1\rangle = \begin{pmatrix}
     0    \\
      1  
\end{pmatrix}
},
\)
notice that 
\(
\sigma_z |1\rangle = -|1\rangle,
\)
so one can count the number of preceding bits that contribute to the overall phase shift due to fermionic bit exchange involving the $\gamma$th qubit with tensor product operator, 
 \(
 \sigma_z^{\otimes \gamma-1} |\psi \rangle = (-1)^{N_\gamma}| \psi\rangle.
\)
The phase factor is determined by the number of bit crossings 
\(
N_\gamma = \sum_{k=1}^{\gamma-1}n_k
\)
in the state $|\psi\rangle$
and where the boolean number variables are $n_k\in [ 0,1]$.
Hence, an annihilation operator is decomposed into a tensor product known as the Jordan-Wigner transformation \cite{JordanWigner1928}
\begin{equation}
\label{new_annihilation_operator_gamma}
a_\gamma =\sigma_z^{\otimes \gamma-1} \otimes\, a\otimes \bm{1}^{\otimes Q-\gamma}
\end{equation}
for integer $\gamma \in [1,Q]$ and here the singleton operator is $a=\left.\frac{1}{2}\middle(\sigma_x + i \sigma_y\right)$, where $\sigma_i$ for $i=x,y,z$ are the Pauli matrices.
See page 17 of \cite{fetter-71} for the usual development for determining $N_\gamma$.
(\ref{new_annihilation_operator_gamma}) and 
its transpose, the creation operator $a^\dagger_\gamma=a_\gamma^\text{\tiny T}$, satisfy the anti-commutation relations
\begin{eqnarray}
\label{fermionic_anti_commutation_relations}
\{   a_\gamma,   a^\dagger_\beta\} & =&  \delta_{\gamma\beta},
\quad
\{   a_\gamma,   a_\beta\}  =   0,
\quad
\{   a^\dagger_\gamma,   a^\dagger_\beta\}  =  0.
\quad
\end{eqnarray}

The 
hermitian generator of a quantum gate
  can be analytically expressed in terms of qubit creation and annihilation operators.  A novel generator that is
manifestly hermitian is the following
\begin{equation}
\label{generalized_interchanger_generator_reparameterized_2}
\begin{split}
{\cal E}_{\Delta\alpha\gamma}
&=
d^{-1}  \Big[
    -A^2  \, n_\alpha
-
 A^{-2} \, n_{\gamma }
\\
&
+
i \, 
\left(
e^{i\xi}a^\dagger_\alpha a_{\gamma} 
-
e^{-i\xi} a^\dagger_{\gamma }   a_\alpha 
\right)
+d \,(\Delta -1)  n_\alpha  n_{\gamma}
\Big],
\end{split}
\end{equation}
where $d = -A^2 -A^{-2}$ is real, and $\xi$ is an internal e-bit phase angle.  The parameter $\Delta$ is boolean, and it allows one to select between fermionic ($\Delta=1$) or bose ($\Delta=0$) statistics of the modeled quantum particles.

The coefficients in (\ref{generalized_interchanger_generator_reparameterized_2}) can be parameterized by a real angle $\mu$: ${\cal E}_{\Delta\alpha\gamma}={\cal E}_{\Delta\alpha\gamma}(\mu)$ with $A^2=-\frac{\cos\mu+1}{\sin\mu}$ and $d = 2\csc\mu$.
 The   quantum logic gate
 generated by ${\cal E}_{\Delta \alpha\gamma}$ is
\begin{equation}
\label{conservative_quantum_logic_gate}
e^{i\zeta\, {\cal E}_{\Delta \alpha\gamma}} = \mathbf{1}^{\otimes \text{\tiny \it Q}}  + (e^{i\zeta} -1) {\cal E}_{\Delta \alpha\gamma}.
\end{equation}
%

% Vertical gate convention
%
%	P3				P4
%
%	V3   	RV3		LV4	V4
%	V1	RV1		LV2	V2
%
%	P1				P2
%

\def\VGate{
\xy
0;/r0.5pc/:
(0,-0.25)*{}="V1";
(0,0.25)*{}="V3";
(1,0)+"V1"*{}="RV1";
(1,0)+"V3"*{}="RV3";
(4.9,0)+"RV3"*{}="LV4";
(4.9,0)+"RV1"*{}="LV2";
(0, 2)+"V3"*{}="P3";
(0,-2)+"V1"*{} ="P1";
"RV1"; "P3" **\crv{"V1"}  \POS?(.35)*{\hole}="x" ; 
"RV3"; "x" **\crv{ (0.25,0.12) } ;
"x"; "P1" **\crv{ (0,-0.25) } ;
%Adjoint matter
"RV1"; "LV2" **\dir{~};
"RV3"; "LV4" **\dir{~};
(1,0)+"LV4"*{}="V4";
(1,0)+"LV2"*{}="V2";
(0, 2)+"V4"*{}="P4";
(0,-2)+"V2"*{} ="P2";
"LV4"; "P4" **\crv{"V4"};
"LV2"; "P2" **\crv{"V2"};
\endxy
}
\def\VGateAdj{
\xy
0;/r0.5pc/:
(0,-0.25)*{}="V1";
(0,0.25)*{}="V3";
(1,0)+"V1"*{}="RV1";
(1,0)+"V3"*{}="RV3";
(4.9,0)+"RV3"*{}="LV4";
(4.9,0)+"RV1"*{}="LV2";
(0, 2)+"V3"*{}="P3";
(0,-2)+"V1"*{} ="P1";
"RV3"; "P1" **\crv{"V1"}  \POS?(.3)*{\hole}="x" ; 
"RV1"; "x" **\crv{ (0.25,0.12) } ;
"x"; "P3" **\crv{ (0,0.25) } ;
%Adjoint matter
"RV1"; "LV2" **\dir{~};
"RV3"; "LV4" **\dir{~};
(1,0)+"LV4"*{}="V4";
(1,0)+"LV2"*{}="V2";
(0, 2)+"V4"*{}="P4";
(0,-2)+"V2"*{} ="P2";
"LV4"; "P4" **\crv{"V4"};
"LV2"; "P2" **\crv{"V2"};
\endxy
}

% Usage" \YepezScatter{ <arrow direction P1> }{ <arrow direction P2> }{ <arrow direction P3> }{ <arrow direction P4> }{ <size> }
\UseComputerModernTips
\def\YepezScatter#1#2#3#4#5{   % Recommend: #1 = 0.5
\mathord{\xybox{0;/r#5pc/:
(0,-0.25)*{}="V1";
(0,0.25)*{}="V3";
(0.5,0)+"V1"*{}="RV1";
(0.5,0)+"V3"*{}="RV3";
(5,0)+"RV3"*{}="LV4";
(5,0)+"RV1"*{}="LV2";
(0, 2)+"V3"*{}="P3";
(0,-2)+"V1"*{} ="P1";
%\hover~{"V3"}{"RV3"}{"V1"}{"RV1"};
%"V3"; "P3" **\dir{-};
%"V1"; "P1" **\dir{-};
(0.3,0)+"RV1"; "P3" **\crv{(-0.3,-.75)+"V1"}  \POS?(.55)*{\hole}="x" ;   ?(0.85)*\dir{#3}; 
(0.25,0.1)+"RV3"; "x" **\crv{(.2,0)+ "V1" } ;
"x"; "P1" **\crv{ (0,-0.25) } ;  ?(0.7)*\dir{#1}; 
%Adjoint matter
"RV1"; "LV2" **\dir{~}; 
"RV3"; "LV4" **\dir{~};
(0.5,0)+"LV4"*{}="V4";
(0.5,0)+"LV2"*{}="V2";
(0, 2)+"V4"*{}="P4";
(0,-2)+"V2"*{} ="P2";
(-0.4,-0.15)+"LV4"; "P4" **\crv{"V4"}; ?(0.85)*\dir{#4}; 
(-0.4,-0.15)+"LV2"; "P2" **\crv{"V2"}; ?(0.7)*\dir{#2}; 
}}
}
\def\YepezScatterAdj#1#2#3#4#5{   % Recommend: #1 = 0.5
\mathord{\xybox{0;/r#5pc/:
(0,-0.25)*{}="V1";
(0,0.25)*{}="V3";
(0.5,0)+"V1"*{}="RV1";
(0.5,0)+"V3"*{}="RV3";
(5,0)+"RV3"*{}="LV4";
(5,0)+"RV1"*{}="LV2";
(0, 2)+"V3"*{}="P3";
(0,-2)+"V1"*{} ="P1";
(0.2,0)+"RV3"; "P1" **\crv{"V1"}  \POS?(.4)*{\hole}="x" ; ?(0.85)*\dir{#3};
"x"; "P3" **\crv{ (0,0.25) } ; ?(0.7)*\dir{#1}; 
%Adjoint matter
"RV1"; "LV2" **\dir{~}; 
"RV3"; "LV4" **\dir{~};
(0.5,0)+"LV4"*{}="V4";
(0.5,0)+"LV2"*{}="V2";
(0, 2)+"V4"*{}="P4";
(0,-2)+"V2"*{} ="P2";
(-0.4,-0.15)+"LV4"; "P4" **\crv{"V4"}; ?(0.85)*\dir{#4}; 
(-0.4,-0.15)+"LV2"; "P2" **\crv{"V2"}; ?(0.7)*\dir{#2}; 
}}
}
The state evolution (\ref{basic_conservative_quantum_gate}) by the quantum logic gate (\ref{conservative_quantum_logic_gate}) can be understood as scattering between two qubits
\begin{subequations}
\label{tree_level_quantum_gate_evolution_and_inverse}
\begin{eqnarray}
\label{tree_level_quantum_gate_evolution}
|\psi'\rangle= e^{i\zeta\, {\cal E}_{\Delta \alpha\gamma}}|\psi\rangle\quad
&\Longleftrightarrow&
\xy
0;/r0.5pc/:
(14,0)*+{\YepezScatter{}{}{}{}{0.4}};
% input qubits
(12,3)*{\text{\scriptsize $|q_\alpha\rangle$}};
(17,3)*{\text{\scriptsize $|q_\gamma\rangle$}};
% output qubits
(12,-3)*{\text{\scriptsize $|q'_\alpha\rangle$}};
(17,-3)*{\text{\scriptsize $|q'_\gamma\rangle$}};
\endxy
\\
\label{tree_level_adjoint_quantum_gate_evolution}
|\psi\rangle= e^{-i\zeta\, {\cal E}_{\Delta \alpha\gamma}}|\psi'\rangle\quad
& \Longleftrightarrow&
\xy
0;/r0.5pc/:
(14,0)*+{\YepezScatterAdj{}{}{}{}{0.4}};
% input qubits
(12,3)*{\text{\scriptsize $|q_\alpha\rangle$}};
(17,3)*{\text{\scriptsize $|q_\gamma\rangle$}};
% output qubits
(12,-3)*{\text{\scriptsize $|q'_\alpha\rangle$}};
(17,-3)*{\text{\scriptsize $|q'_\gamma\rangle$}};
\endxy
\end{eqnarray}
\end{subequations}
where the ``gauge field'' that couples the external qubit world-lines is represented by an internal double wavy line (or ribbon).  The external lines either over-cross or under-cross and are assigned $+1$ and $-1$ multiplying the action, {\it i.e.}  $\pm \zeta {\cal E}_{\Delta}$. This sign disambiguates between a quantum gate and its adjoint, respectively, as shown in (\ref{tree_level_quantum_gate_evolution}) and (\ref{tree_level_adjoint_quantum_gate_evolution}).  Let us denote a qubit graphically $|q_\alpha\rangle \equiv u_\alpha \uparrow + d_\alpha \downarrow$, with complex amplitudes constrained by conservation of probability $|u_\alpha|^2+|d_\alpha|^2=1$. 

Starting, for example, with a separable input state $|\psi\rangle = |q_\alpha\rangle|q_\gamma\rangle$, a scattering diagram is a quantum superposition of four oriented graphs
\vspace{-0.1in}
\begin{equation}
\label{quantum_scattering_example_separable_1}
\begin{split}
\xy
(0,-0.5)*+{\YepezScatter{}{}{}{}{0.4}};
% input qubits
(-3,6)*{\text{\scriptsize $|q_\alpha\rangle$}};
(6,6)*{\text{\scriptsize $|q_\gamma\rangle$}};
\endxy 
& =  
u_\alpha u_\gamma
\xy
(0,0.5)*+{\YepezScatter{<}{<}{>}{>}{0.4}};
\endxy 
+ 
u_\alpha d_\gamma
\xy
(0,0.5)*+{\YepezScatter{>}{<}{>}{<}{0.4}};
\endxy 
\\
& +
d_\alpha u_\gamma
\xy
(0,0.5)*+{\YepezScatter{<}{>}{<}{>}{0.4}};
\endxy 
+ 
d_\alpha d_\gamma
\xy
(0,0.5)*+{\YepezScatter{>}{>}{<}{<}{0.4}};
\endxy .
\end{split}
\end{equation}
Each oriented scattering graph can be reduced to a quantum superposition of classical graphs, or just a single classical graph, as the case may be.  
There are four {\bf quantum skein relations} representing dynamics generated by (\ref{generalized_interchanger_generator_reparameterized_2})
%
% Usage:  \ClassicalGateDD { <arrow direction at x> } { <arrow direction at y> } { <arrow direction at z> }\knotcrossing { <mirror> } { <size> }
\def\ClassicalGate#1#2#3#4#5#6{\mathord{\xybox{0;/r#6pc/:#4#5<#1|#2>#3}}}
\vspace{-0.25in}
\begin{subequations}
\label{quantum_skein_relations}
\begin{eqnarray}
%(a)
\label{quantum_skein_relations_a}
\xy
(0,0.5)*+{\YepezScatter{<}{<}{>}{>}{0.4}};
\endxy & = & 
\xy
(0,.5)*+{\ClassicalGate{|}{>}{>}\vuntwist{-}{1.5}};
\endxy 
\\
%(b)
%\nonumber
\label{quantum_skein_relations_b}
\xy
(0,0.5)*+{\YepezScatter{>}{<}{>}{<}{0.4}};
\endxy
&=&
\text{\scriptsize $\frac{-A^2-A^{-2} e^{i\zeta}}{d}$}
\xy
(-0.1,0)*{\ClassicalGate{<}{>}{|}\vuntwist{}{1.5}};
\endxy 
%\\
%&&
\text{\scriptsize $\ + \frac{i e^{-i\xi}\left(e^{i\zeta}-1\right)}{d}$ }
\xy
(0,0)*{\ClassicalGate{>}{<}{|}\vtwistneg{}{1.5}};
\endxy
\quad \ \ \
\\
%(c)
%\nonumber
\label{quantum_skein_relations_c}
\xy
(0,0.5)*+{\YepezScatter{<}{>}{<}{>}{0.4}};
\endxy
 & = &  
\text{\scriptsize $\frac{-A^2 e^{i\zeta}-A^{-2} }{d}$}
\xy
(0,.5)*+{\ClassicalGate{>}{<}{|}\vuntwist{}{1.5}};
\endxy 
%\\
%&&
\text{\scriptsize $- \frac{i e^{i\xi}\left(e^{i\zeta}-1\right)}{d}$}
\xy
(0,.5)*+{\ClassicalGate{>}{<}{|}\vtwist{}{1.5}};
\endxy
\\
%(d)
\label{quantum_skein_relations_d}
\xy
(0,0.5)*+{\YepezScatter{>}{>}{<}{<}{0.4}};
\endxy
 & = &  
\text{\scriptsize $1 + \left(e^{i\zeta} - 1\right) \Delta$}
\xy
(0,.5)*+{\ClassicalGate{|}{>}{>}\vuntwist{}{1.5}};
\endxy.
\end{eqnarray}
\end{subequations}
These are the quantum analog of (\ref{skein_relations}).  
Adjoint quantum skein relations are obtained simply by taking $\zeta\rightarrow -\zeta$ in the amplitudes in the diagrams in (\ref{quantum_skein_relations}).
All superbraids can be reduced to a quantum superposition of classical braids.  The closure of a superbraid forms a superlink.  Hence, a superlink can be reduced to a quantum superposition of classical links, and consequently, for each superlink one can compute a superlink invariant, for example a superposition of Jones polynomials.

In the context of quantum information dynamics, a physical interpretation of the parameter $A$ can be rendered as follows.
If the strands in $L$ are considered closed spacetime histories of $Q$ qubits (e.g. qubit states evolving in a quantum circuit with closed-loop feedback), then the L.H.S. of (\ref{quantum_skein_relations}) represent a trajectory configuration within a piece of the superlink where entanglement is generated by a qubit-qubit coupling that occurs at a quantum-gate ({\it i.e.} generalized crossing point).  For the 1-body cases (\ref{quantum_skein_relations_b}) and (\ref{quantum_skein_relations_c}), the R.H.S.  represents classical alternatives in quantum superposition:  $d^{-1}(-A^2 - A^{-2}e^{i\zeta})$ is the amplitude for no interaction (non-swapping of qubit states) whereas the amplitude of a {\sc swap} interaction (interchanging of qubit states) goes as $d^{-1}(e^{i\zeta}-1)$.

As an example of reducing a superbraid, let us recursively apply (\ref{quantum_skein_relations}) to prove an obvious evolution identity: the composition of a quantum gate with its adjoint is the identity operator, {\it i.e.} $U U^\dagger=1$.  For simplicity,  we start with $|q_\alpha\rangle = \uparrow$ and $|q_\gamma\rangle=\downarrow$, so the initially oriented superbraid is reduced to a superposition of classical braids as follows:
\begin{widetext}
\begin{eqnarray}
%(a)
\xy
0;/r1pc/:
(0.2,1.5)*+{\text{\scriptsize \it U}};
(0,0.9)*+{\YepezScatter{}{}{>}{<}{0.4}};
(0.2,-0.2)*+{\text{\scriptsize \it U${}^\dagger$}};
(0.05,-0.9)*+{\YepezScatterAdj{}{}{}{}{0.4}};
\endxy
&\stackrel{(\ref{quantum_skein_relations_b})}{=}&
\text{\scriptsize $\frac{-A^2-A^{-2} e^{i\zeta}}{d}$}
\xy
0;/r1pc/:
(0.08,0.7)*+{\ClassicalGate{<}{>}{|}\vuntwist{}{1.65}};
(0,-0.7)*+{\YepezScatterAdj{}{}{}{}{0.28}};
\endxy
\text{\scriptsize $+ \frac{i e^{-i\xi}\left(e^{i\zeta}-1\right)}{d}$}
\xy
0;/r1pc/:
(0.16,0.7)*+{\ClassicalGate{>}{<}{|}\vtwistneg{}{1.65}};
(0,-0.7)*+{\YepezScatterAdj{}{}{}{}{0.28}};
\endxy
\end{eqnarray}
\vspace{-0.35in}
\begin{eqnarray}
%(b)
\nonumber
\stackrel{(\ref{quantum_skein_relations_b})^{\dagger}}{=}
&&
\hspace{-0.15in} 
\text{\scriptsize $\frac{-A^2-A^{-2} e^{i\zeta}}{d}$} \
\text{\scriptsize $\frac{-A^2-A^{-2} e^{-i\zeta}}{d}$}
\xy
0;/r1pc/:
(0,0.65)*+{\ClassicalGate{<}{|}{|}\vuntwist{}{1.25}};
(0.13,-0.65)*+{\ClassicalGate{|}{>}{|}\vuntwist{}{1.25}};
\endxy
+
\text{\scriptsize $\frac{-A^2-A^{-2} e^{i\zeta}}{d}$} \
\text{\scriptsize $ \frac{i e^{-i\xi}\left(e^{-i\zeta}-1\right)}{d}$}
\xy
0;/r1pc/:
(0,0.65)*+{\ClassicalGate{<}{|}{|}\vuntwist{}{1.25}};
(0.15,-0.65)*+{\ClassicalGate{>}{|}{|}\vtwistneg{}{1.25}};
\endxy
\text{\scriptsize $+ \frac{i e^{-i\xi}\left(e^{i\zeta}-1\right)}{d}$}
\xy
0;/r1pc/:
(0.16,0.7)*+{\ClassicalGate{>}{<}{|}\vtwistneg{}{1.65}};
(0,-0.7)*+{\YepezScatterAdj{}{}{}{}{0.28}};
\endxy
\\
%(c)
\stackrel{(\ref{quantum_skein_relations_c})^{\dagger}}{=}
&&
\hspace{-0.15in} 
\nonumber
\text{\scriptsize $\frac{A^4+A^{-4} +2\cos\zeta}{d^2}$} 
\xy
0;/r1pc/:
(0,0)*+{\ClassicalGate{<}{>}{|}\vuntwist{}{1.25}};
\endxy
\text{\scriptsize $- \frac{i e^{-i\xi}}{d^2}$}
\text{\scriptsize $\left(A^2+A^{-2} e^{i\zeta}\middle)\middle(e^{-i\zeta}-1\right)$}
\xy
0;/r1pc/:
(0.0,0)*+{\ClassicalGate{>}{<}{|}\vtwistneg{}{1.25}};
\endxy
\text{\scriptsize $- \frac{i e^{-i\xi}}{d^2}$}
\text{\scriptsize $\left(A^2 e^{-i\zeta}+A^{-2} \middle)\middle(e^{i\zeta}-1\right)$}
\xy
0;/r1pc/:
(0,0.65)*+{\ClassicalGate{>}{|}{|}\vtwistneg{}{1.25}};
(0.0,-0.65)*+{\ClassicalGate{|}{<}{|}\vuntwist{}{1.25}};
\endxy
\text{\scriptsize $+ \frac{\left(e^{-i\zeta}-1\right)\left(e^{i\zeta}-1\right)}{d^2}$}
\xy
0;/r1pc/:
(0.1,0.65)*+{\ClassicalGate{>}{|}{|}\vtwistneg{}{1.25}};
(-0.1,-0.65)*+{\ClassicalGate{|}{<}{|}\vtwist{}{1.25}};
\endxy
\\
%(d)
\hspace{-0.15in} 
\stackrel{(\ref{braid_identity_1})}{=}
&&
\hspace{-0.15in} 
\nonumber
\text{\scriptsize $\frac{A^4+A^{-4} +2\cos\zeta}{d^2}$} 
\xy
0;/r1pc/:
(0,0)*+{\ClassicalGate{<}{>}{|}\vuntwist{}{1.25}};
\endxy
\text{\scriptsize $ \ - \frac{i e^{-i\xi}}{d^2}$}
\text{\scriptsize $\left(A^2e^{-i\zeta}-A^{-2} e^{i\zeta}-A^2+A^{-2}\right)$}
\left(
\xy
0;/r1pc/:
(0.0,0)*+{\ClassicalGate{>}{<}{|}\vtwistneg{}{1.25}};
\endxy
-
\xy
0;/r1pc/:
(0,0)*+{\ClassicalGate{>}{<}{|}\vtwistneg{}{1.25}};
\endxy
\right)
\text{\scriptsize $\ + \ \frac{2-2\cos\zeta}{d^2}$}
\xy
0;/r1pc/:
(0,0)*+{\ClassicalGate{<}{>}{|}\vuntwist{}{1.25}};
\endxy
\\
\nonumber
%(e)
=
&&
\xy
0;/r1pc/:
(0,0)*+{\ClassicalGate{<}{>}{|}\vuntwist{}{1.25}};
\endxy.
\end{eqnarray}
\end{widetext}
It is easy to verify the same result occurs for inputs $|q_\alpha\rangle = \downarrow$ and $|q_\gamma\rangle=\uparrow$. Furthermore, the identity trivially follows for $|q_\alpha\rangle = \uparrow$ and $|q_\gamma\rangle=\uparrow$ and for $|q_\alpha\rangle = \downarrow$ and $|q_\gamma\rangle=\downarrow$ since $\Delta$ is boolean.

With adjacent indices, {\it e.g.} $\gamma=\alpha+ 1$ in (\ref{conservative_quantum_logic_gate}), we need write the first index only ({\it i.e.} suppress the second indice),
\(
{\cal E}_{\Delta\alpha}  \equiv  {\cal E}_{\Delta\alpha,\alpha+ 1} .
\)
Using this compressed notation, 
(\ref{generalized_interchanger_generator_reparameterized_2}) satisfies the following {\bf quantum Temperley-Lieb algebra}
\begin{subequations}
\label{Yepez_algebra_Y_Q}
\begin{eqnarray}
   {\cal E}_{\Delta\alpha}^2 &=&    {\cal E}_{\Delta\alpha}, \ \ \qquad \alpha=1,2,\dots, Q-1\qquad
   \\
   \nonumber
   {\cal E}_{\Delta\alpha} {\cal E}_{\Delta\alpha\pm1} {\cal E}_{\Delta\alpha}&-&   {\cal E}_{\Delta\alpha\pm1} {\cal E}_{\Delta\alpha} {\cal E}_{\Delta\alpha\pm1} = 
   \\
\label{Yepez_algebra_Y_Q_b}
&+&      
 d^{-2} {\cal E}_{\Delta\alpha} -d^{-2} {\cal E}_{\Delta\alpha\pm1}
   \\
   {\cal E}_{\Delta\alpha} {\cal E}_{\Delta\beta} &=&{\cal E}_{\Delta\beta}  {\cal E}_{\Delta\alpha}, \quad |\alpha-\beta| \ge 2.
      \end{eqnarray}
\end{subequations}
To help understand this algebra, we may write (\ref{Yepez_algebra_Y_Q_b}) as follows
\begin{subequations}
\label{interleaving_separation}
\begin{eqnarray}
   {\cal E}_{\Delta\alpha} {\cal E}_{\Delta\alpha+1} {\cal E}_{\Delta\alpha}-d^{-2} {\cal E}_{\Delta\alpha}&=& d^{-2} X_{\alpha,\alpha+ 1}\, 
   \\
    {\cal E}_{\Delta\alpha+1} {\cal E}_{\Delta\alpha} {\cal E}_{\Delta\alpha+1} -d^{-2} {\cal E}_{\Delta\alpha+1} &=& d^{-2} Y_{\alpha,\alpha+ 1} ,\quad
\end{eqnarray}
\end{subequations}
where $X_{\alpha,\alpha+ 1}$ and $Y_{\alpha,\alpha+ 1}$ are introduced solely for the purpose of separating  (\ref{Yepez_algebra_Y_Q_b})  into two equations.  For (\ref{interleaving_separation}) to be equivalent to (\ref{Yepez_algebra_Y_Q_b}), one must demonstrate that $X_{\alpha,\alpha+ 1}=Y_{\alpha,\alpha+ 1}$.
 Inserting (\ref{generalized_interchanger_generator_reparameterized_2}) into the L.H.S. of (\ref{interleaving_separation}), after considerable ladder operator algebra, 
one finds that the difference of the R.H.S. of (\ref{interleaving_separation}) is
\begin{eqnarray}
\nonumber
X_{\alpha,\alpha+ 1}-Y_{\alpha,\alpha+ 1} & = &  \Delta(\Delta -1)\big[ (A^4-A^{-4})n_\alpha n_{\alpha+1} n_{\alpha+2}
\\
& -& A^4 n_\alpha n_{\alpha+1} + A^{-4}n_{\alpha+1} n_{\alpha+2}\big] ,
\end{eqnarray}
 vanishing for boolean $\Delta$.  Thus, (\ref{Yepez_algebra_Y_Q_b}) follows from (\ref{generalized_interchanger_generator_reparameterized_2}).

One finds $X$ and $Y$ are proportional to $\Delta$, so a remarkable reduction of  (\ref{Yepez_algebra_Y_Q}) occurs for the $\Delta =0$ case:
\begin{subequations}
\begin{eqnarray}
   {\cal E}_{0\alpha}^2 &=&    {\cal E}_{0\alpha}, \ \qquad \alpha=1,2,\dots, Q-1\qquad
   \\
   {\cal E}_{0\alpha} {\cal E}_{0\alpha\pm1} {\cal E}_{0\alpha}&\stackrel{(\ref{interleaving_separation})}{=}&d^{-2} {\cal E}_{0\alpha}
   \\
   {\cal E}_{0\alpha} {\cal E}_{0\beta} &=&{\cal E}_{0\beta}  {\cal E}_{0\alpha}, \quad |\alpha-\beta| \ge 2.
      \end{eqnarray}
\end{subequations}
This is the Temperley-Lieb algebra over a system of $Q$ qubits (TL${}_Q$).  
Thus, entangled bosonic states generated by $ {\cal E}_{0\alpha}$ are isomorphic to links generated by ${\cal E}_{0\alpha}$.
So (\ref{Yepez_algebra_Y_Q}) is a generalization of TL${}_Q$.
We now consider the generalized braid that it generates:  a superbraid.

A general {\bf superbraid operator} is an amalgamation of both a classical braid operator and a quantum gate
\begin{equation}
\label{superbraid_operator_defined}
b^\text{s}_{\Delta\alpha\beta} \equiv A \, e^{z\, {\cal E}_{\Delta \alpha\beta}},
\end{equation}
where $A$ and $z$ are complex parameters. 
(\ref{superbraid_operator_defined}) can be applied to any two qubits, $\alpha$ and $\beta$, in a system of qubits ({\it i.e.} we do not impose a restriction to the adjacency case when $\beta=\alpha+1$).
  (\ref{superbraid_operator_defined}) can be written in several different ways, each way useful in its own right.

Letting $z\equiv i\zeta + \ln \tau$, the superbraid operator has the following exponential form
\begin{subequations}
\begin{eqnarray}
%\label{ }
b^\text{s}_{\Delta\alpha\beta}&\equiv& \tau^{4} \,e^{(i\zeta + \ln \tau)\, {\cal E}_{\Delta \alpha \beta}}
=
 \tau^{4} \left( e^{i\zeta} \tau\right)^{{\cal E}_{\Delta \alpha\beta}},\qquad
\end{eqnarray}
\end{subequations}
where $\tau^4\equiv A$.  The superbraid operator can be written linearly in its generator
\begin{subequations}
\begin{eqnarray}
b^\text{s}_{\Delta\alpha\beta} 
 \label{superbraid_operator_A_d_theta_form}
 &=&
A 
 \left[\mathbf{1}_\text{\tiny \it Q}  + (A^{-4} \,e^{i\zeta} -1) {\cal E}_{\Delta \alpha\beta}\right]
\\
&=&
A \,\mathbf{1}_\text{\tiny \it Q}  +A^{-1} d 
\left( 
\frac{1-e^{i\zeta}\tau}{1+\tau}
\right)
{\cal E}_{\Delta \alpha\beta}. \qquad
\end{eqnarray}
\end{subequations}
A non-trivial classical limit of quantum logic gates represented as (\ref{conservative_quantum_logic_gate}) occurs at $\zeta=\pi$  ({\sc swap} operator).  
Consequently, the superbraid operator in product form is
\begin{subequations}
\begin{eqnarray}
%\label{ }
b^\text{s}_{\Delta\alpha\beta} &\equiv& \tau^{4} \,e^{ (\ln \tau+i\pi)\, {\cal E}_{\Delta \alpha \beta}}\,e^{(i\zeta-i\pi) \, {\cal E}_{\Delta \alpha \beta}}
 \\
 &=&
 \label{supebraid_opeator_amalgamation_form}
b_{\Delta\alpha\beta} \,e^{i(\zeta-\pi)\, {\cal E}_{\Delta \alpha \beta}},
\end{eqnarray}
\end{subequations}
where $b_{\Delta\alpha\beta}=\tau^{4} \,e^{ (\ln \tau+i\pi)\, {\cal E}_{\Delta \alpha \beta}}$ is the conventional braid operator. (\ref{supebraid_opeator_amalgamation_form}) is useful for comprehending the physical behavior of the superbraid operator.  It classically braids world lines $\alpha$ and $\beta$ and quantum mechanically entangles these world lines according to the deficit angle $\zeta-\pi$.

The superbraid group is defined by 
\begin{subequations}
\label{superbraid_algebra}
\begin{eqnarray}
\label{superbraid_algebra_1}
b^\text{s}_\alpha \,b^\text{s}_\beta & = & b^\text{s}_\beta\,b^\text{s}_\alpha ,\ \  \, \quad\text{for}\quad |\alpha-\beta|>1 \quad\\
\label{superbraid_algebra_2}
b^\text{s}_\alpha\, b^\text{s}_{\alpha+ 1}\, b^\text{s}_\alpha 
+ \gamma  \,b^\text{s}_{\alpha}
& = &  b^\text{s}_{\alpha+1} b^\text{s}_{\alpha} b^\text{s}_{\alpha+1} 
+ \gamma  \,b^\text{s}_{\alpha+1},
\\
\nonumber
&&\  \quad\qquad \quad\text{for}\quad 1\le \alpha <Q,
\qquad
\end{eqnarray}
\end{subequations}
where $\gamma$ is a constant that depends on the representation.   For (\ref{generalized_interchanger_generator_reparameterized_2}),  we have  
\(
\gamma = 
\left(
A^4 + A^{-4} e^{i\zeta}
\middle)
\middle(
1+ e^{i\zeta}
\right) A^{-2} d^{-2}.
\)

 In the classical limit $\zeta=\pi$, the superbraid operator reduces to the classical braid operator, $b_\alpha\equiv b^\text{s}_\alpha(\pi, \tau)$, and   (\ref{superbraid_algebra}) reduces to the Artin braid group
\begin{subequations}
\label{Braid_algebra}
\begin{eqnarray}
\label{Braid_algebra_1}
b_\alpha \,b_\beta & = & b_\beta\,b_\alpha ,\ \  \, \qquad \quad\text{for}\quad |\alpha-\beta|>1 \\
\label{Braid_algebra_2}
b_\alpha\, b_{\alpha+ 1}\, b_\alpha & = &  b_{\alpha+1} b_{\alpha} b_{\alpha+1},  \quad\text{for}\quad 1\le \alpha <Q.
\qquad
\end{eqnarray}
\end{subequations}
(\ref{Braid_algebra}) follows from (\ref{superbraid_algebra}) because $\gamma=0$ for $\zeta=\pi$.  Also, in this classical limit, (\ref{superbraid_operator_A_d_theta_form}) reduces to the  braid operator
\(
b_\alpha = A \, \mathbf{1}_\text{\tiny \it Q} + A^{-1} d\, {\cal E}_{\Delta \alpha \beta},  
\)
for $\alpha=1,2,\dots, Q-1$.
After some ladder operator algebra, one finds that
\begin{equation}
\begin{split}
& b_\alpha  b_{\alpha+1} b_\alpha  - b_{\alpha+1} b_{\alpha} b_{\alpha+1} = 
\\
& A^{-1}(A^4-A^{-4}) d^{-2} \Delta(\Delta-1) \, (1-n_\alpha)  n_{\alpha+1}  n_{\alpha+2},
\end{split}
\end{equation}
where $n_\alpha\equiv a^\dagger_\alpha a_\alpha$. 
 Since $\Delta$ is boolean, the R.H.S. vanishes, and this is just (\ref{Braid_algebra_2}).

\bibliographystyle{apsrev}

\end{document}